# Imaging half-unit-cell Cooper-pair density waves in monolayer 1T′-MoTe$_2$

Fang-Jun Cheng[1,*], Cong-Cong Lou[1,*], Ai-Xi Chen[2], Li-Xuan Wei[1], Yu Liu[1], Bo-Yuan Deng[2], Fangsen Li[2,†], Ziqiang Wang[3,†], Qi-Kun Xue[1,4,5,6,7,†], Xu-Cun Ma[1,4], Can-Li Song[1,4,†]

[1]State Key Laboratory of Low-Dimensional Quantum Physics, Department of Physics, Tsinghua University, Beijing 100084, China

[2]Vacuum Interconnected Nanotech Workstation, Suzhou Institute of Nano-Tech and Nano-Bionics, Chinese Academy of Sciences, Suzhou 215123, China

[3]Department of Physics, Boston College, Chestnut Hill, MA 0246, USA

[4]Frontier Science Center for Quantum Information, Beijing 100084, China

[5]Shenzhen Institute for Quantum Science and Engineering and Department of Physics, Southern University of Science and Technology, Shenzhen 518055, China

[6]Hefei National Laboratory, Heifei 230088, China

[7]Beijing Academy of Quantum Information Sciences, Beijing 100193, China

Unconventional superconductors that spontaneously break space-group symmetries of their underlying crystal lattice are distinguished by spatial modulations of the superconducting order parameter. These states have recently captured significant attention in various strongly correlated materials, where the breaking of translational or intra-unit-cell symmetries results in the emergence of pair density waves with wavelengths extending across one or multiple unit cells. Here, we employ a spectroscopic-imaging scanning tunneling microscope to identify robust half-unit-cell modulations of the superconducting gap magnitude, coherence strength and subgap states in tunable 1T′-MoTe$_2$ monolayer. The modulations are oriented perpendicular to the zigzag Mo chains and coexist with three-unit-cell primary pair density wave modulations along the Mo chains, both of which attenuate with increasing temperature and the application of external magnetic fields. Importantly, we find that the superconductivity modulations are strongly linked to unconventional electron pairing mechanisms, which significantly deviate from the Cooper pairing observed in traditional Bardeen-Cooper-Schrieffer superconductors. Our results advance the knowledge of Cooper-pair density modulations and the intricate interplay with other symmetry-breaking states in strongly correlated superconductors.

*These authors contributed equally to this work.

†To whom correspondence should be addressed. Email: clsong07@mail.tsinghua.edu.cn, fsli2015@sinano.ac.cn, ziqiang.wang@bc.edu, qkxue@mail.tsinghua.edu.cn

Unconventional superconductors have captivated attentions for decades due to their complex and often extraordinary properties arising from the intricate interplay between electronic correlations and symmetry-breaking phenomena[1,2]. Among the most exotic aspects of these materials is their ability to spontaneously break translational symmetries of the underlying crystal lattice, leading to nontrivial spatial modulations of the superconducting order parameter[3-21]. The periodic modulations at wave vector $Q_P$ are associated with the emergence of pair density waves (PDWs)[5], a novel state of matter characterized by Cooper pairing with finite center-of-mass momentum $\hbar Q_P$ (where $\hbar$ is Planck's constant $h$ divided by $2\pi$). This ground state can induce various vestigial orders such as charge density waves (CDWs), charge-$4e$ and charge-$6e$ superconductivity[22-24], which stands in stark contrast to the conventional Bardeen-Cooper-Schrieffer (BCS) superconductivity with zero momentum electron pairing and the magnetic field induced Fulde-Ferrell-Larkin-Ovchinnikov (FFLO) states[25,26]. An increasing number of experiments have offered spectroscopic and transport evidences for PDWs, either as the primary state or as a subsidiary order accompanying CDWs, in numerous correlated compounds including the cuprates[6-12], transition metal dichalcogenides[13], UTe$_2$[15,16], kagome materials[14,19,20], and iron-based superconductors[17,18]. Nevertheless, the wavelength of these PDWs, associated with $Q_P$ of the finite momentum Cooper pairs, extends over several unit cells of the underlying lattice constant. Remarkably, single-unit-cell PDWs have been recently observed from the breaking of intra-unit-cell point-group symmetry or an unconventional $\eta$-pairing in iron dichalcogenides[27-30]. Despite these advances, the microscopic mechanism of PDWs and their interplay with other symmetry-breaking states remains to be elucidated.

Monolayer (ML) MoTe$_2$ in its distorted octahedral 1T′ structure, driven by a Peierls distortion (Fig. 1a, inset), has emerged as a promising platform for exploring unconventional superconductivity and associated symmetry-breaking states[31,32]. Compared to its bulk counterparts, which have a superconducting transition temperature ($T_c$) of around 0.1 K[33], ML 1T′-MoTe$_2$ demonstrated significantly enhanced superconducting properties, whether exfoliated from the crystal[34] or epitaxially grown on various substrates[35,36]. Additionally, this material's multi-band electronic structure and tunability through ambipolar doping render it a potential candidate for exploring the interplay among unconventional superconductivity, electronic correlations, and symmetry breaking states[36-39]. In this study, we apply spectroscopic imaging scanning tunneling microscopy (SI-STM) to reveal coexistent PDWs and CDWs at distinct wave vectors in moderately electron-doped ML 1T′-MoTe$_2$ films. Interestingly, the superconductivity exhibits identical spatial variations across distinct Te sublattices, resulting in half-unit-cell PDW modulations oriented perpendicular to the zigzag Mo chains.

High-quality ML 1T′-MoTe$_2$ films were epitaxially prepared on graphitized SiC(0001) substrates using a state-of-the-arts molecular beam epitaxy (MBE) under Te-rich conditions (see Methods for more details)[36]. Through interfacial engineering, ML 1T′-MoTe$_2$ exhibits sizable tunability in its unique electronic structure and superconducting properties across various samples and regions (Supplementary Note 1). This tunability is illustrated in Supplementary Fig. 1 by measuring sample (and space)-dependent differential conductance spectra $g(\mathbf{r}, V) \equiv dI/dV(\mathbf{r}, V)$ as a function of the energy $E = eV$ at spatial coordinate $\mathbf{r}$ ($e$ is the elementary charge). In moderately electron-doped ML 1T′-MoTe$_2$ films, an electron pocket is revealed to emerge at the $X$ point of the Brillouin zone (BZ) boundary by studying energy-resolved quasiparticle interference patterns (Extended Data Fig. 1). Hence, the Fermi surface (FS) comprises a hole pocket centered at the $\Gamma$ point, and three electron pockets at the $X$ point and along the $\Gamma Y$ line, as illustrated in Fig. 1a. This situation is highly reminiscent of iron-based superconductors[40] and multiple inter(intra)-band/pocket electron pairing channels become possible. Among these, intra-band interactions within the electron pockets are particularly interesting since they can potentially pair electrons up on the same side of the FS. These novel pairing states, discussed in terms of the Amperean pairing in unconventional superconductors[21,41], carry finite momenta $\hbar Q_P^a = (\pm 2\pi/a, 0)\hbar$ and $\hbar Q_P^b \approx (0, \pm 2\pi/3b)\hbar$ in ML 1T′-MoTe$_2$. Notably, $Q_P^a$ connects the reciprocal lattice vector $Q_{Bragg}$ along the $q_a$ axis and is governed by Umklapp interactions. Albeit typically weak, this term may be substantially enhanced in strength when the FS is proximate to the edges of the BZ in multi-band compounds, just like the electron pockets at $X$ in 1T′-MoTe$_2$.

Figure 1b shows a representative topographic image $T(\mathbf{r}, V)$ of the moderately electron-doped ML 1T′-MoTe$_2$ films with atomic resolution, acquired within a field of view (FOV) of 28 nm × 28 nm. In addition to atomic Bragg peaks, the Fourier transform (FT) of $T(\mathbf{r}, V)$ reveals distinct Fourier peaks at $Q_C^a \approx (\pm \pi/a, 0)$ and $Q_C^b \approx (0, \pm 2\pi/5b)$ (Fig. 1c, circled in blue). These peaks correspond to the emergence of bidirectional CDWs with a 2 × 5 supercell relative to the unit cell, contrasting with the unidirectional CDW observed in lightly doped ML 1T′-MoTe$_2$ films[36]. Given that the wave vectors $Q_C^{a(b)}$ approximately connect the electron and hole pockets, our findings underscore the importance of inter-band interactions between these pockets in stabilizing the CDWs. Notably, the Fourier peaks at $Q_C^{a(b)}$ exhibit subtle splitting even after correction of the lattice distortions of the $T(\boldsymbol{q})$ image using the Lawleer-Fujita algorithm[42]. This splitting may most likely originate from an imperfect nesting between the shape-distinctive electron and hole pockets, thereby giving rise to fluctuating rather than static CDWs.

To make the visualization of the fluctuating CDWs more clearly, a Gaussian filter is applied to the $T(\boldsymbol{q})$ data. By selectively retaining only the Fourier components around $Q_C^a$ or $Q_C^b$ (detailed in the Method section) and then performing an inverse Fourier transform, the spatial modulations of CDWs corresponding to these wave vectors are readily revealed. In Fig. 1d,e and Supplementary Fig. 2, we show real part $T(\boldsymbol{r}, Q_C^{a(b)})$ of the inverse Fourier transforms, revealing apparent spatial inhomogeneity and topological defects. These perturbations disturb the long-range charge order, a hallmark of fluctuating CDWs. Moreover, we observe the CDWs are highly sensitive to the interfacial engineering, with fluctuations at $Q_C^b$ becoming pronounced when the electron doping level increases (Supplementary Fig. 3). Accompanying the enhanced fluctuations at $Q_C^b$, the superconducting energy gap structure experiences a profound transformation as well, changing from a predominant gap at 1.27 meV to 1.85 meV (Supplementary Fig. 1b). In Fig. 1f,g, we depict the temperature and external magnetic field evolutions of the superconducting gap in the moderately electron-doped ML 1T′-MoTe$_2$ films. Based on these data, we extract $T_c \approx 6.0$ K and upper critical field $B_{c2} \approx 4.0$ T. Interestingly, above these thresholds, the $dI/dV$ curves display apparent spectral dips near $E_F$ with an energy scale considerably smaller than the CDW gap $\Delta_C \sim 67$ meV (see the inset of Supplementary Fig. 1a). This is reminiscent of the pseudogap-like phenomenon observed in many correlated superconductors[1,14].

The most striking finding arises from mapping the local superconducting gap magnitude $|\Delta(\boldsymbol{r})|$, defined as half the distance between the two coherence peaks in the spatially-resolved d$I$/d$V$ spectra, as illustrated in Fig. 2a. Clearly, $\Delta(\boldsymbol{r})$ exhibits prominent modulations along the crystalline $a$ direction with a uniform half-unit-cell ($a/2$, ~ 3.2 Å) wavelength, corresponding to a unique wave vector of $2Q_P^a = (4\pi/a, 0)$. By aligning $\Delta(\boldsymbol{r})$ with the simultaneously acquired topography $T(\boldsymbol{r})$, we reveal that the gap maxima are indistinguishably situated on the two distinctly buckled Te sublattices at the bottom layer of 1T′-MoTe$_2$ (Supplementary Fig. 4). Illustrated in Fig. 2b is the FT of $|\Delta(\boldsymbol{r})|$, which reveals pronounced Fourier peaks at $\pm 2Q_P^a$ and $Q_P^b \approx (0, \pm 2\pi/3b)$ (outlined by black squares). The $Q_P^b$ peaks signify additional three-unit-cell (~ 3$b$) modulations of the $\Delta(\boldsymbol{r})$ along the $b$ direction, which, albeit weak, are discernible in Fig. 2a. Figure 2c plots line cuts of the FT through the two distinct FT peaks along the two orthogonal $q_a$ and $q_b$ axes and underscores the sharpness of the $2Q_P^a$ and $Q_P^b$ peaks, each appearing as nearly one single pixel in width. These modulations are robust across various sample regions (Extended Data Fig. 2), indicating the emergence of bidirectional PDWs at zero magnetic field. The spatial $\Delta(r)$ modulations at $2Q_P^a$ ($Q_P^b$) are estimated to be 0.035 meV (0.03 meV) in magnitude (Methods and Extended Data Fig. 3). These results, particularly the unprecedented discovery of half-unit-cell Cooper-pair density wave modulations, represent a significant breakthrough in our interfacial engineering of the ML 1T′-MoTe$_2$ films.

Figure 2d shows a map of the integrated superconducting coherence peak intensities $C(r)$, as defined in Fig. 1f, which further corroborates the bidirectional spatial modulations of superconductivity. Linecut d$I$/d$V$ spectra measured along the crystalline *a* and *b* directions are depicted in Fig. 2e and 2f, respectively. These spectra feature particle-hole symmetric spatial modulations both near the superconducting coherence peaks and within the energy gaps $\Delta(r)$. To elucidate the energy dependency of these modulations, we present the energy-varying $g(r, E)$ maps and linecuts of the corresponding $g(q, E)$ maps along the $q_a$ axis in Extended Data Fig. 4. It is immediately evident that the bidirectional modulations predominantly manifest within the superconducting energy gaps and attenuate at higher energies. Critically, these modulations are discernible at low energies down to $E_F$ ($V = 0$), indicating the presence of residual zero-energy states. This phenomenon aligns with the rounded bottom of the superconducting gaps at 0.4 K (Fig. 1f,g) and the finite-momentum electron pairing scenario[19], but contrasts with the conventional *s*-wave superconductivity. Importantly, the wave vectors $2Q_P^a$ and $Q_P^b$ for the Cooper-pair density modulations are completely distinct from those ($Q_C^{a(b)}$) for the CDWs (Fig. 1d,e). A coupling of the fluctuating CDW orders with a uniform superconducting order parameter ($\Delta_0 e^{i\theta_0}$) potentially induces subsidiary modulations of $\Delta(r)$ at $Q_C^{a(b)}$. However, these modulations are negligibly weak (Fig. 2b,c and Extended Data Fig. 2b,c) within the limit of our measurement resolutions. Collectively, our observations elucidate the origin of the superconductivity modulations at $2Q_P^a$ and $Q_P^b$ from primary PDW instabilities in the moderately electron-doped ML 1T′-MoTe$_2$.

The superconductivity modulations at $Q_P^b$ can be attributed to the Amperean pairing associated with the electron pockets along the $\Gamma Y$ line (Fig. 1a)[21,41], because these pockets are justly centered at a momentum of $\sim Q_P^b/2$[32,36-38]. Such a pairing emerges at the expense of the unidirectional PDWs with a wave vector $Q_C^b \approx (0, \pm 2\pi/5b)$, which were previously identified in the lightly doped case[36]. Similarly, the Amperean pairing from the electron pockets at $X$ might account for the modest $\Delta(r)$ modulations at $Q_P^a$ seen in Fig. 2c and Extended

Data Fig. 2c. However, the spatial modulations of superconductivity at $2Q_P^a$ present a particularly intriguing conundrum, as they manifest on the half-unit-cell scale with greater intensity than those at $Q_P^a$. A plausible cause involves a breaking of time-reversal symmetry in the superconducting state (Supplementary Note 2), which has been extensively discussed in various multi-band superconductors[20,43,44]. Considering the one-by-one correspondence between the $\Delta(r)$ maxima and the distinct Te sublattices (Fig. 2a and Supplementary Fig. 4), however, a more reasonable explanation for the $2Q_P^a$ superconductivity modulations must go beyond four-band models that contain Mo atoms and only one Te sub lattice (red-circled in Fig. 1a) in the unit cell[45]. With equivalent involvement of both Te sublattices, electrons may preferentially pair between the distinct Te sublattices rather than within the same ones ($\eta$-pairing)[46,47]. As a result, the sublattice structure naturally induces spatial modulations in $\Delta(r)$, characterized by the half-unit-cell periodicity along the *a* direction, or equivalently, the wave vector of $2Q_P^a$ in ML 1T′-MoTe$_2$ films. The reason why $\Delta(r)$ peaks at the bottom Te atoms may be most probably associated with the interfacial electron transfer from the underlying substrates, which will induce an out-of-plane electric field in the ML 1T′-MoTe$_2$ films. This electric field undoubtedly breaks the inversion symmetry between the bottom and top Te lattices (Fig. 1a), thereby causing the peaked $\Delta(r)$ to occur at the electron-accumulating Te atoms at the bottom layer.

Regardless of the microscopic mechanism underlying these PDWs, a distinguishing hallmark that sets these PDWs apart from a uniform superconductivity is their capacity to induce various vestigial orders such as CDWs ($\rho_Q$)[3-5]. These charge modulations can be readily visualized by mapping the subgap states $g(r, V)$. Figure 3a-h illustrate a series of zero-bias conductance (ZBC) maps (i.e. $g(r, E = 0)$) measured at $T = 0.4$ K, 4.2, 10 K and 15 K within the identical FOV as Fig. 1b along with their Fourier transforms, while Extended Data Fig. 5 shows the energy-dependent $g(q, E)$ maps at 0.4 K. We observe no CDW order associated with $\rho_{P_a-P_b} \propto (\Delta_P^a \Delta_P^{b*} + \Delta_{-P}^b \Delta_{-P}^{a*})$ within the superconducting gap (Fig. 3e-h and Extended Data Fig. 5). This implies a phase difference of $\pi/2$ between the two PDW components along the *a* and *b* directions, $\Delta_P^a(r)$ and $\Delta_P^b(r)$, a characteristic analogous to those observed in cuprate superconductors[5,8,9]. Instead, we identify two prominent CDWs at two wave vectors $2Q_P^a$ and $Q_P^b$, which primarily stem from coupling of the bidirectional PDWs to the uniform superconducting order parameter, namely $\rho_{P_a} \propto \Delta_P^a \Delta_0 e^{-i\theta_0} + \Delta_{-P}^a \Delta_0 e^{i\theta_0} \propto \cos(2Q_P^a \cdot r)$ and $\rho_{P_b} \propto \Delta_P^b \Delta_0 e^{-i\theta_0} + \Delta_{-P}^b \Delta_0 e^{i\theta_0} \propto \cos(Q_P^b \cdot r)$. Importantly, the subsidiary CDW orders manifest exclusively within the superconducting energy gaps, as demonstrated in Supplementary Fig. 5.

Despite spatial inhomogeneity in ZBC (Fig. 3a), the subsidiary CDWs exhibit remarkable robustness at 0.4 K irrespective of the spatial variations (Supplementary Fig. 6). As the temperature is elevated, the CDW modulations gradually become weaker (Fig. 3f-h), and the FT intensities at $2Q_P^a$ and $Q_P^b$ display an energy-independent behavior above $T_c$ (Supplementary Fig. 5). Figure 3i plots the CDW intensities calculated from the FT intensity difference between the d$I$/d$V$ maps at $E_F$ and outside the superconducting gaps, specifically $g(2Q_P^a, 0) - g(2Q_P^a, -2\text{ meV})$ and $g(Q_P^b, 0) - g(Q_P^b, -2\text{ meV})$. Evidently, the CDW modulations induced by the bidirectional PDWs approximately follow a mean-field temperature dependence and attenuates significantly above 5.8 K, a temperature around $T_c$. To further highlight the half-unit-cell PDW modulations, we compare linecuts of the energy-dependent $g(\boldsymbol{q}, E = eV)$, derived from the $g(\boldsymbol{r}, V)$ maps measured over the entire 28 × 28 nm FOV, through Γ along the $q_a$ axis below (Fig. 3j, 0.4 K) and above $T_c$ (Fig. 3k, 15 K). The signals of $2Q_P^a$ predominantly manifest within the superconducting gaps at 0.4 K and vanish above $T_c$. This provides compelling evidence for the existence of half-unit-cell PDWs in ML 1T′-MoTe$_2$.

Next, we focus on the half-unit-cell PDW modulations and examine the magnetic field dependence of the accompanying CDWs ($\rho_{P_a}$). As elaborated in Fig. 4a, the CDW order at $2Q_P^a$ is yet observable within the superconducting gaps below $B_{c2}$, but exhibit considerable attenuation as the superconductivity is suppressed by a larger magnetic field of 6 T. It is noteworthy that residual spatial modulations at $2Q_P^a$ persist even as the magnetic field of 6 T exceeds the upper critical field $B_{c2}$. Simultaneously, the CDWs at the wave vector $Q_C^a$ ≈ (±π/$a$, 0), as well as quasiparticle interferences stemming from the inter(intra)-electron-pocket scatterings, re-appear in the subgap state maps $g(\boldsymbol{r}, V)$ (Fig. 4b). The unique coexistence of $2Q_P^a$ and $Q_C^a$ CDWs provides us an opportunity to explore the intricate interplay between the PDWs and CDWs in real space. To achieve this, we analyze the distributions of the modulation amplitudes $A_P^a(\boldsymbol{r}, 2Q_P^a)$ and $A_C^a(\boldsymbol{r}, Q_C^a)$ from the ZBC map at 6 T using the standard lock-in approach (Methods). A visual comparison indicates a negative relationship between the $A_P^a(\boldsymbol{r}, 2Q_P^a)$ and $A_C^a(\boldsymbol{r}, Q_C^a)$ maps (Fig. 4c,d), having a correlation coefficient of approximately -0.49. This suggests that both of the half-unit-cell PDW modulations at $2Q_P^a$ and CDWs at $Q_C^a$ are intimately linked to the emergent electron pockets at the $X$ point of the BZ. They naturally compete with each other in their spectroscopic manifestations.

Finally, we comment on the pseudogap-like phenomenon observed in Fig. 1f,g. The pseudogap has the same energy scale as the superconducting gap $\Delta(\boldsymbol{r})$, within which the bidirectional PDWs uniquely develop (Extended Data Fig. 4 and 5). This differs markedly from the 2 ×5 CDWs, which manifest at a considerably higher energy scale than $\Delta(r)$ (Supplementary Fig. 2 and 3). These findings suggest a compelling possibility that the pseudogap might originate from fluctuating PDWs above $T_c$ and $B_{c2}$, a concept extensively explored in recent studies on cuprate and other unconventional superconductors[5,14,21,48]. This proposal is substantiated by revealing a synchronized evolution between the pseudogap and PDW states, both vanishing far above $T_c$ (Fig. 1f and 3h) but persisting at 0.4 K and above $B_{c2}$ (Fig. 1g and 4b). According to the Amperean pairing scenario[21], the pseudogap phase can be viewed as a novel pairing state that automatically develops when the long-range PDW order is destroyed by phase fluctuations. From this perspective, our experimental findings of pseudogap-like characteristics serves as additional evidence for the primary PDWs in ML 1T′-MoTe$_2$.

Irrespective of the multi-band interactions that facilitate Amperean pairing within the electron pockets and the microscopic origins of the sublattice electron pairing, our experimental findings undoubtedly reveal bidirectional Cooper-pair density wave modulations involving multiple intra(inter)-band pairing channels in the moderately electron-doped ML 1T′-MoTe$_2$ films. Importantly, the unprecedented identification of half-unit-cell modulations in the superconducting order parameters, which involves an unconventional sublattice electron pairing, provides profound insights into the interatomic variations of superconducting pairing, and their intricate interplay with the translational symmetry-breaking CDW orders. These findings pave the way for further theoretical and experimental investigations on the microscopic mechanisms underpinning the recently extensively explored PDW states and their interplay with other charge ordered states in the unconventional superconductors. Our ability to induce and manipulate these unusual states through an interfacial engineering and doping strategy in ML 1T′-MoTe$_2$ opens new avenues for designing novel materials with tailored electronic properties, potentially leading to significant advancements in uncovering hidden quantum phases and elucidating the fundamental physics of unconventional superconductivity in correlated systems.

**Methods**

**Sample growth.** The film growth began with nitrogen-doped 4*H*-SiC(0001) wafers, exhibiting resistivities between 0.015 Ω cm and 0.028 Ω cm. These wafers were graphitized by heating to 1400°C for 15 minutes under ultra-high vacuum conditions, producing graphene substrates that vary in thickness from single-layer to multiplayers. This spatial variability in graphene thickness offers a unique opportunity to finely tune the properties of the overlayers grown atop them. High-purity molybdenum (99.95%) and tellurium (99.9999%) were then co-deposited on these graphene/SiC substrates. The molybdenum atoms were evaporated from a single-pocket electron beam evaporator, and the tellurium from a standard Knudsen diffusion cell, ensuring an extremely high Te/Mo flux ratio exceeding 20. This high ratio is crucial to compensate for the volatility of tellurium during growth, as excess tellurium molecules desorbed at a moderate substrate temperature of 300 °C. This specific temperature was selected to inhibit the formation of the more stable 2*H*-$MoTe_2$ phase, favoring the controlled growth of large-area MLs of 1T′-$MoTe_2$ films. Lower substrate temperatures tended to result in less continuous ML 1T′-$MoTe_2$ films, often interspersed with bilayer $MoTe_2$. To ensure optimal surface cleanliness for our STM measurements, the samples were annealed under a tellurium atmosphere at 300 °C for 20 minutes before being transferred to the STM apparatus.

***In-situ* STM measurement.** STM measurements were conducted using a state-of-the-art Unisoku USM 1300 system, interconnected with an MBE chamber for *in-situ* sample preparation. Both chambers maintain a base pressure below $2.0 \times 10^{-10}$ Torr. The system allows for the application of a maximum magnetic field of 8 T perpendicular and 2 T parallel to the sample surface. Polycrystalline PtIr tips, conditioned through *e*-beam bombardment in MBE chamber and calibrated on crystalline Ag/Si(111) films, were employed throughout the experiments. STM topographies $T(r)$ were acquired in constant current mode, while tunneling d$I$/d$V$ spectra and conductance maps $g(\mathbf{r}, V)$ were recorded using a high-precision lock-in technique with a small a.c. modulation voltage at a frequency of 983 Hz.

**Two-dimensional lock-in analysis.** To extract amplitude maps from STM topographies or spectroscopic maps, we utilized a standard two-dimensional lock-in technique[42]. For example, an STM topography $T(\mathbf{r})$ can be regarded as a superposition of periodic signals, encompassing atomic lattice and other modulations. By applying a Fourier transform of the $T(\mathbf{r})$, one can effectively distinguish potential modulation signals. This methodology is expressed as

$$T(\mathbf{r}) = \sum_Q A_Q(\mathbf{r}) e^{i\mathbf{Q}\cdot\mathbf{r}} \tag{1}$$

$$T(\mathbf{q}) = \sum_Q A_Q(\mathbf{q} - \mathbf{Q}), \tag{2}$$

where $A_Q(\mathbf{q} - \mathbf{Q})$ denotes the amplitude distribution function of a modulation at wave vector $\mathbf{Q}$. For an ideal plane wave, the amplitude $A_Q(\mathbf{q} - \mathbf{Q})$ would concentrate at a single pixel. However, various factors such as thermal fluctuations, tip drift, lattice strain and defects can cause inhomogeneous distribution, broadening the single peak from one pixel to several. To extract the modulation signal at wave vector $\mathbf{Q}$, we utilize a standard lock-in method which involves multiplying $T(\mathbf{q})$ (Fourier transform of $T(\mathbf{r})$) by a Gaussian packet centered at $\mathbf{Q}$, followed by an inverse Fourier transform

$$T_Q(\mathbf{r}) = F^{-1}\left[T(\mathbf{q}) \frac{1}{\sqrt{2\pi}\sigma_Q} \exp\left(-\frac{(\mathbf{q}-\mathbf{Q})^2}{2\sigma_Q^2}\right)\right]. \tag{3}$$

The cutoff length $\sigma_Q$ of the Gaussian packet in the $\mathbf{q}$ space should satisfy

$$\lambda_Q < \sigma_Q \ll \delta_Q, \tag{4}$$

where $\lambda_Q$ represents the characteristic broadening length of $A_Q(\mathbf{q} - \mathbf{Q})$ and $\delta_Q$ is the distance between $\mathbf{Q}$ and the nearest wave vector around it. The extracted single wave vector map $T_Q(\mathbf{r})$ can be described as

$$T_Q(\mathbf{r}) = A(\mathbf{r}, \mathbf{Q})\exp[-i(\mathbf{Q}\cdot\mathbf{r} + \phi(\mathbf{r}, \mathbf{Q}))]. \tag{5}$$

As thus, we can derive the amplitude map $A(\mathbf{r}, \mathbf{Q})$ and phase map $\phi(\mathbf{r}, \mathbf{Q})$ from the inverse Fourier transform

$$A(\mathbf{r}, \mathbf{Q}) = |T_Q(\mathbf{r})| \tag{6}$$

$$\phi(\mathbf{r}, \mathbf{Q}) = -\mathbf{Q}\cdot\mathbf{r} + i\ln\frac{T_Q(\mathbf{r})}{A(\mathbf{r}, \mathbf{Q})}. \tag{7}$$

In addition, one can also obtain the real part of the inverse Fourier transform

$$T(\mathbf{r}, \mathbf{Q}) = A(\mathbf{r}, \mathbf{Q})\cos(\phi(\mathbf{r}, \mathbf{Q})), \tag{8}$$

which includes both the magnitude and phase information of the modulations at wave vector $\mathbf{Q}$.

To elucidate the effect of filter size on the inverse Fourier transform, we depict the dependence of $T(r, Q_C^a)$ and $T(r, Q_C^b)$ on the real-space cutoff length $\sigma_r = 1/\sigma_Q$ in Fig. 1d,e and Supplementary Fig. 2. Here, the $\sigma_r$ values span from 20 Å to 40 Å, with the minimum value slightly exceeding the CDW wavelength of ~ $5b$ along the $a$ direction. For larger $\sigma_r$, the condition $\lambda_Q < \sigma_Q$ cannot be met. Notably, the key features and the distribution of topological defects in the filtered $T(r, Q_C^{(a,b)})$ images remain essentially unchanged to great extent. This demonstrates the accuracy of our lock-in analysis and the robustness of fluctuation CDWs in moderately electron-doped ML 1T′-MoTe$_2$ films. For uniformity, a filter size of 25 Å is used for the lock-in analysis in Fig. 4c,d and Supplementary Fig. 3.

**Extraction of spatial modulations in $\Delta(r)$.** To extract the superconducting gap modulations along the $a$ and $b$ directions, the values of $|\Delta(r)|$ in Fig. 2a are spatially averaged over the $b$ and $a$ axes, respectively. This technique ensures a precise characterization of the bidirectional PDWs by reducing noise and highlighting $|\Delta(r)|$ modulations along the two directions. The averaged gap magnitudes are shown in Extended Data Fig. 3, which, in addition to displaying periodic modulations with wavelengths of $a/2$ and ~ $3b$, reveal larger-length scale $|\Delta(r)|$ variations attributable to the spatial inhomogeneity in the uniform superconducting gap $\Delta_0$. To quantitatively describe these features, we fit the spatial variations of $|\Delta(r)|$ to a composite function

$$|\Delta(r)| = \Delta_q \cos(qr + \Phi) + A_0 + A_1 r + A_2 r^2 + A_2 r^3 + A_2 r^4, \tag{9}$$

where the first term corresponds to the PDW modulations at $q = (2Q_P^a, Q_P^b)$ and the quartic term accounts for the underlying spatial inhomogeneity in $\Delta_0$. Based on the fitting results illustrated by the light blue lines, the gap modulations at $2Q_P^a$ and $Q_P^b$ are estimated to be $\Delta_{2Q_P^a} = 0.035$ meV and $\Delta_{Q_P^b} \approx 0.030$, respectively.

**Data availability**

All data that support the findings of this study are available from the corresponding authors upon reasonable request.

**Acknowledgments**

We thank C. J. Wu and H. Yao for helpful discussions. The work is financially supported by grants from the National Key Research and Development Program of China (Grant No. 2022YFA1403100), the Natural Science Foundation of China (Grant No. 1214140 and Grant No. 12134008), the Innovation Program for Quantum Science and Technology (2021ZD0302502), Basic Research Development Program of Suzhou

**Author contributions**

C.L.S., F.L., X.C.M. and Q.K.X. conceived and supervised the research. F.J.C., A.X.C., L.X.W. and B.Y.D. grew the samples and performed the STM measurements. C.C.L., F.J.C., C.L.S. and Y.L. analyzed the experimental data and plotted the figures. Z.W. contributed to the theoretical explanation. C.L.S., F.J.C. and C.C.L. wrote the manuscript with comments from all of the authors.

**Competing financial interests**

The authors declare no competing financial interests.

**Additional information**

Supplementary information includes 6 figures and the corresponding figure captions and discussions.

†These authors contributed equally to this work.

*Correspondence and request for materials should be addressed to C.L.S. (email: clsong07@mail.tsinghua.edu.cn), F.L. (email: fsli2015@sinano.ac.cn), Z.W. (email: ziqiang.wang@bc.edu) or Q.K.X. (email: xucunma@mail.tsinghua.edu.cn).

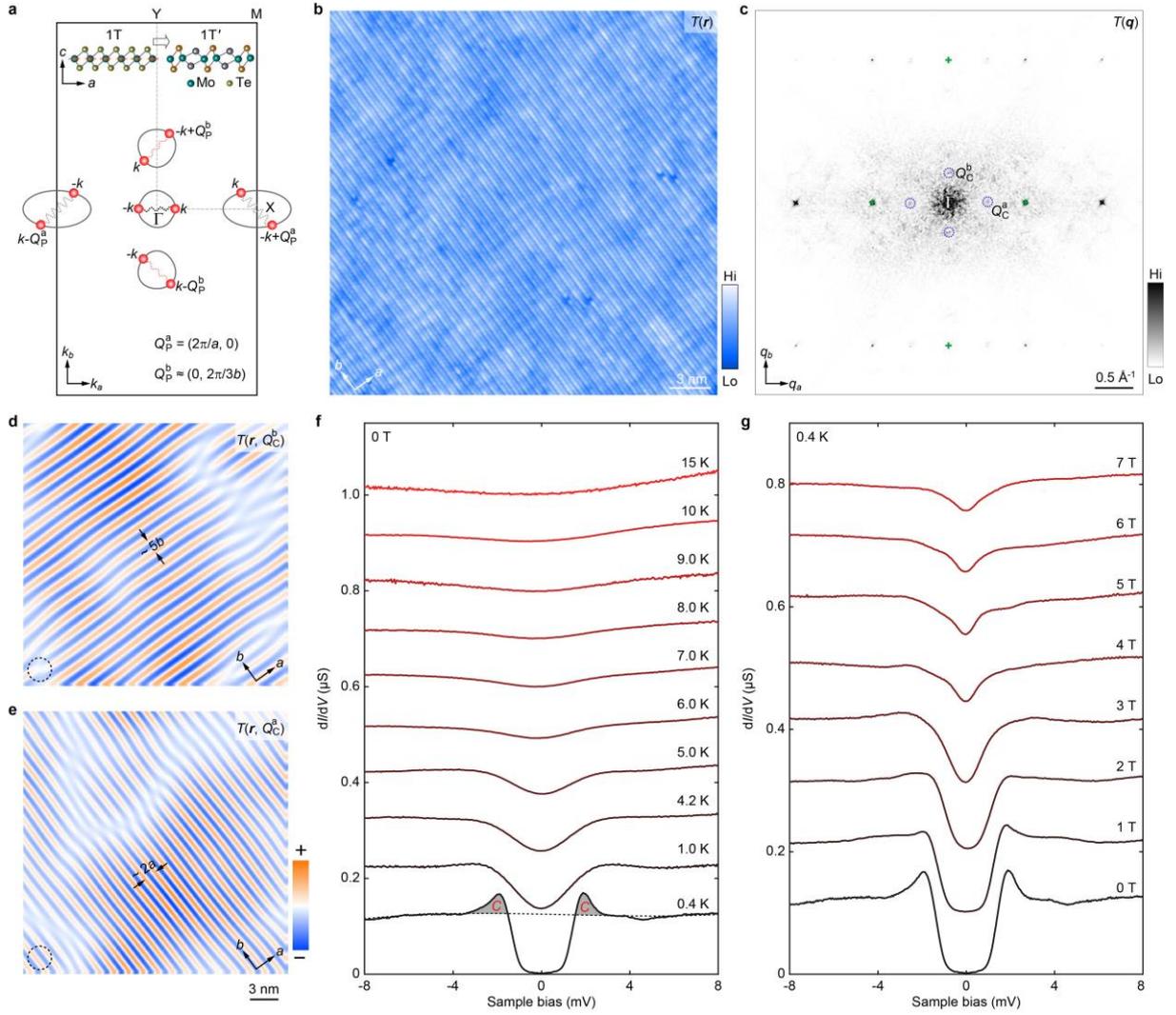

**Figure 1 | Characterization of interfacial electron-doped ML 1T′-MoTe$_2$. a**, Schematic FS showing the multi-band electronic structure in moderately electron-doped ML 1T′-MoTe$_2$. The black rectangle highlights the BZ boundary, while $a(b)$, $k_a(k_b)$ and $q_a(q_b)$ axes are defined to be orthogonal (parallel) to the zigzag Mo chains. Intra-band electron pairing channels are schematically shown in every Fermi pocket. Inset illustrates the Peierls distortion of MoTe$_2$ from the 1T (left) to 1T′ structures (right). In the 1T′ phase, the Te atoms can be divided into two distinct groups (circled in red and blue), with two atoms in each group connected by inversion symmetry. **b**, A 28 nm × 28 nm FOV topography $T(r)$ of ML 1T′-MoTe$_2$ films. Setpoint: $V = 10$ mV, $I = 1$ nA. **c**, Amplitude $T(q)$ derived from a Fourier transform of the drift-corrected $T(r)$, showing bidirectional CDWs at wave vectors $Q_C^a$ and $Q_C^b$. The Bragg peaks at $(\pm 2\pi/a, 0)$ and $(0, \pm 2\pi/b)$ are marked by green crosses, signifying the structural periodicity throughout. **d,e**, Gaussian-filtered $T(r, q)$ images showing spatial modulations exclusively around the CDW wave vectors $q = Q_C^a$ (f) and $Q_C^b$ (g). The filter size of the inverse Fourier transforms is 25 Å, circled in white. **f,g**, Dependences of d$I$/d$V$ spectra on temperature and

magnetic field, respectively, measured at a 10 MΩ junction resistance (setpoint: $V$ = 10 mV, $I$ = 1 nA). The dashed line in **f** represents a sloped background of the d$I$/d$V$ spectrum at 0.4 K, obtained by fitting the raw conductance outside the superconducting gap. Using this as a reference, we define the coherence strength $C$ as the integral over gray-shed areas around the superconducting coherence peaks.

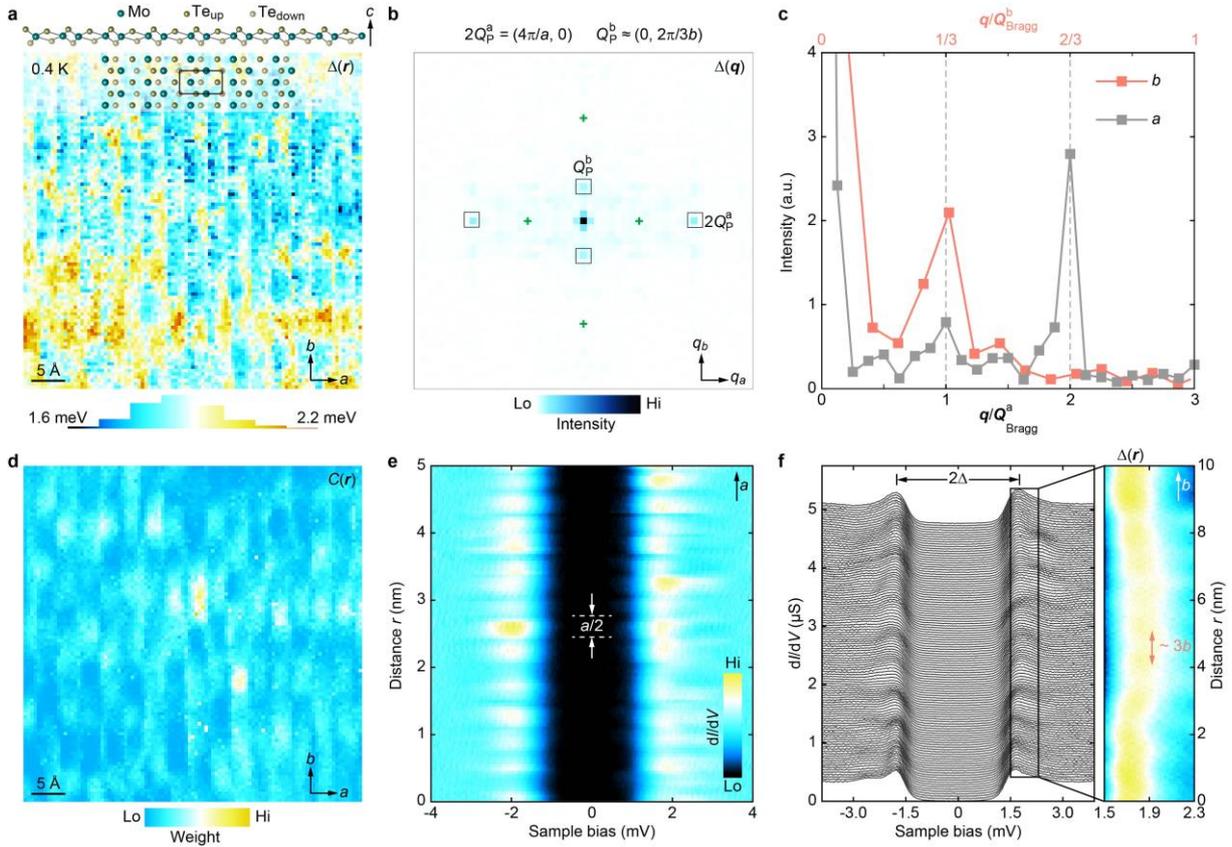

**Figure 2 | Atomic-scale modulations of the superconducting energy gaps. a**, Superconducting gap map |Δ($r$)| exhibiting bidirectional modulations within a 5 nm × 5 nm FOV. Overlaid is the top view of the atom structure of 1T′-MoTe$_2$, with the side view depicted in the top panel. For enhanced clarity, the bottom and top Te atoms are rendered with varying opacities. **b**, Derived |Δ($q$)| from |Δ($r$)| showing two sharp peaks at $2Q_P^a$ and $Q_P^b$ (outlined by black squares). **c**, Line cuts of |Δ($q$)| in **b** along the $q_a$ (gray) and $q_b$ (orange) axes, normalized by the respective Bragg peak wave vectors $Q_{Bragg}^{a(b)}$. **d**, Superconducting coherence strength map $C(r)$ in the same FOV of **a**. **e,f**, Spatially-resolved d$I$/d$V$ spectra measured at equal intervals along the $a$ and $b$ directions, respectively, showing atomic-scale modulations in the superconducting gaps. An intensity map close to the superconducting coherence peaks is inserted in **f** to highlight the superconductivity modulations along the $b$ direction. Setpoint: $V$ = 4 mV, $I$ = 1 nA.

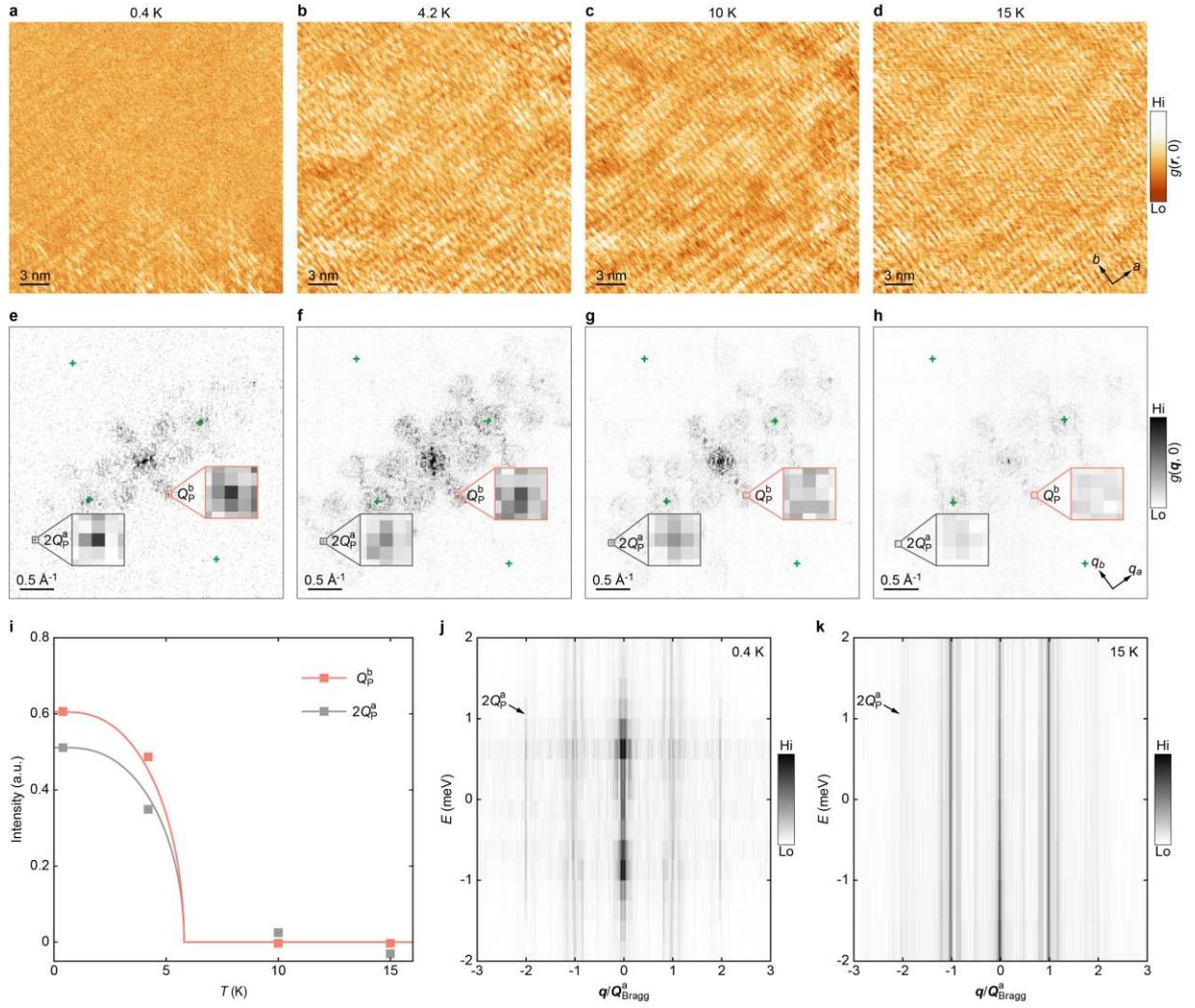

**Figure 3 | Subsidiary CDWs and their temperature dependencies. a-d**, ZBC maps $g(r, E = 0)$ acquired in the same FOV of 28 nm × 28 nm at various temperatures as indicated. Setpoint: $V = 10$ mV, $I = 1$ nA. **e-h**, Derived $g(q, 0)$ from **a-d** showing gradual suppression of the subsidiary CDWs at the wave vectors $2Q_P^a$ and $Q_P^b$, marked by gray and orange squares, respectively. **i**, Temperature dependence of the $g(2Q_P^a, 0)$ and $g(Q_P^b, 0)$ peak intensities, subtracted by the respective values at $E = -2.0$ meV. **j,k**, Maps of line cuts of the energy-dependent $g(q, E)$ through Γ along the $q_a$ axis at temperatures below (**j**) and above $T_c$ (**k**). The $2Q_P^a$ signals are substantially suppressed at 15 K.

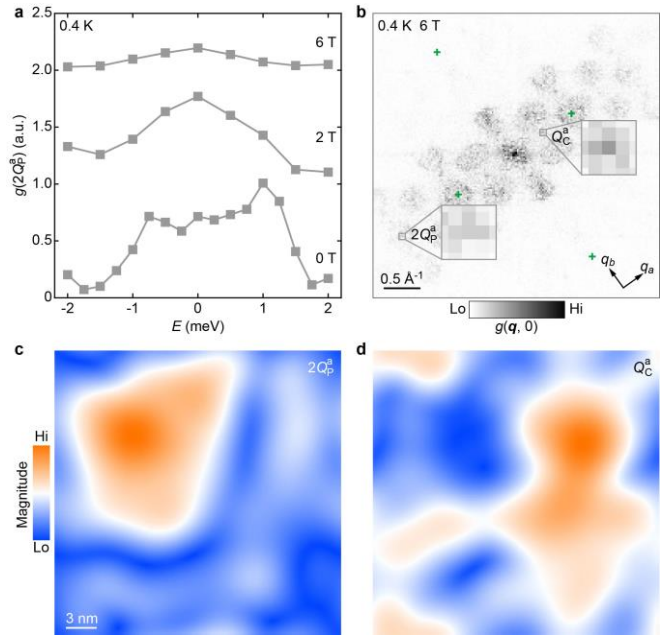

**Figure 4 | Competition of PDW and CDW modulations in magnetic fields. a**, Extracted $g(2Q_P^a, E)$ peak intensities as a function of energy $E$ and external magnetic fields, applied along the $c$ axis. The values at 2 T and 6 T have been vertically offset by 1 and 2 units, respectively, for clarity. **b**, $g(\mathbf{r}, E = 0)$ map measured at $T = 0.4$ K and $B = 6$ T, revealing suppression (re-entrance) of the CDW modulations at $2Q_P^a$ ($Q_C^a$). Setpoint: $V = 10$ mV, $I = 1$ nA. **c,d**, Amplitude maps of the CDW order parameters associated with $2Q_P^a$ and $Q_P^a$ at 6 T, respectively. The two order parameters are inversely correlated with a correlation coefficient of -0.49.

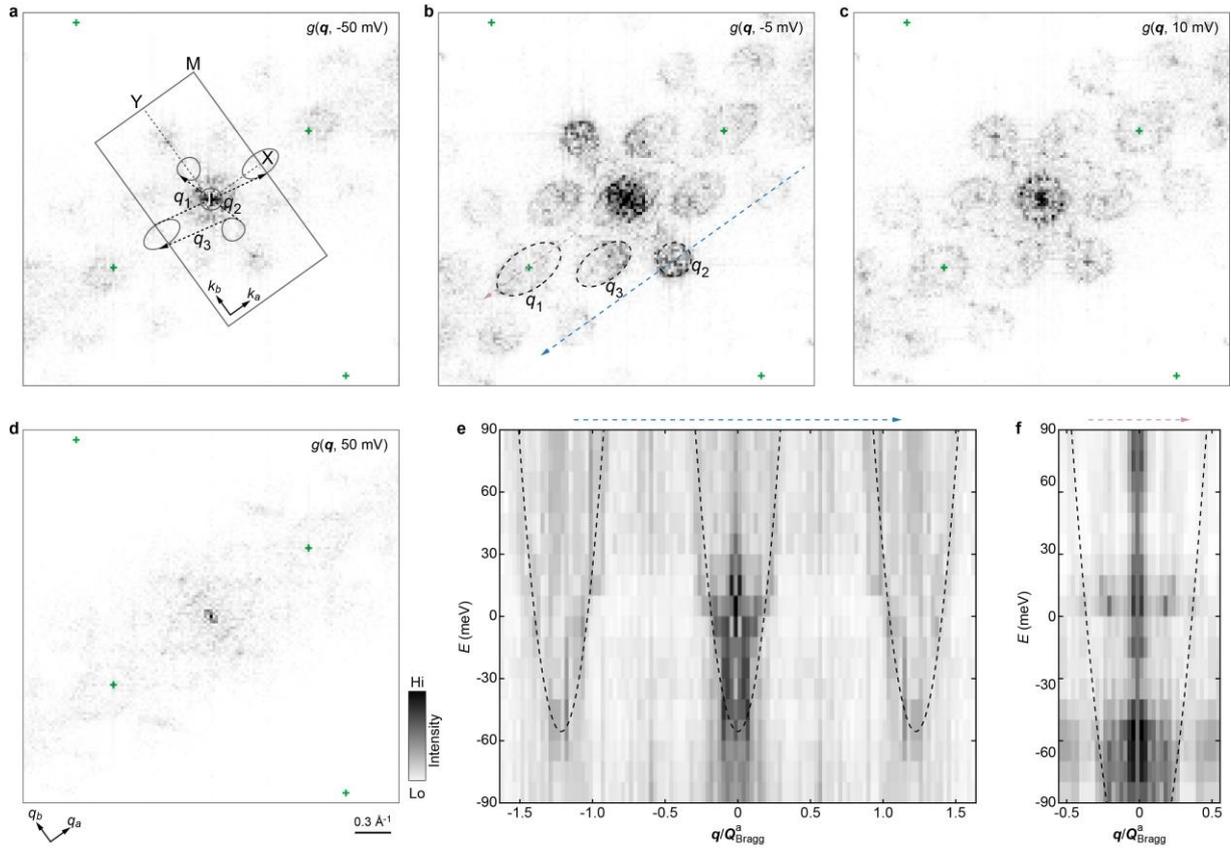

**Extended Data Fig. 1| Evidence for electron pockets at *X*. a-d**, Derived $g(\boldsymbol{q}, E)$ from energy-dependent $g(\boldsymbol{r}, E = eV)$ measurements within the same 28 nm × 28 FOV as Fig. 1b. The black rectangle in **a** delineates the BZ, while the dashed arrows in **a** mark intra (inter)-pocket scatterings among different electron pockets. These scatterings result in three distinctive quasiparticle interference (QPI) patterns, labeled as $q_1$, $q_2$ and $q_3$ and highlighted in **b**. In FS, the emergent electron pockets at the *X* point primarily result from a moderate interfacial electron transfer. **e,f**, Maps of line cuts of the energy-varying $g(\boldsymbol{q}, E)$ intensities along the light blue and magenta arrows marked in **b**, respectively. The scattering wave vector increases with energy. This indicates electron-like band dispersions, highlighted by the parabolic dashes.

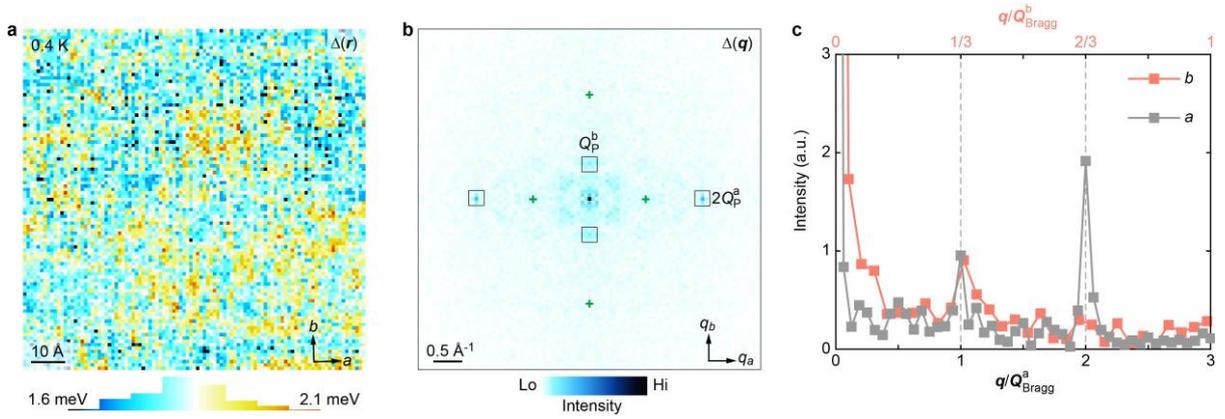

**Extended Data Fig. 2 |. Reproducibility of bidirectional PDW modulations. a**, Mapping superconducting energy gap |Δ(***r***)| in a larger FOV of 10 nm × 10 nm, highlighting consistent atomic-scale modulations in gap magnitude. **b**, |Δ(***q***)| derived from |Δ(***r***)| in **a**, revealing robust Fourier peaks at $2Q_P^a$ and $Q_P^b$ (enclosed by black squares). **c**, Line cuts of |Δ(***q***)| in **b** through Γ along the $q_a$ (gray) and $q_b$ (orange) axes, normalized by the respective Bragg peak wave vectors $Q_{\text{Bragg}}^{a(b)}$. Importantly, the peak intensity of $2Q_P^a$ is twice of that of $Q_P^a$, suggesting more pronounced superconductivity modulations at $2Q_P^a$ compared to $Q_P^b$.

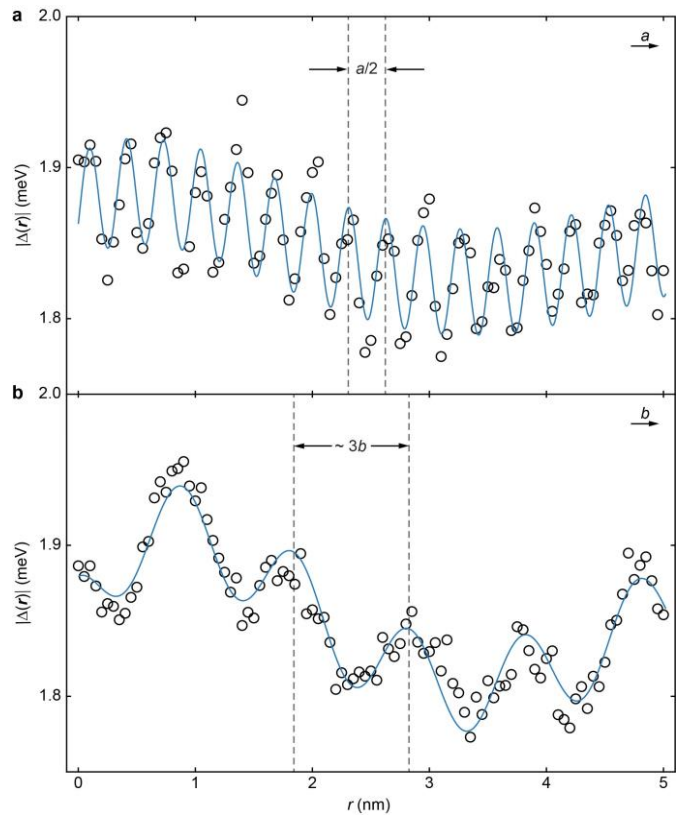

**Extended Data Fig. 3 |. Spatial modulations of superconducting gaps. a,b**, Spatial modulations of |Δ(***r***)| along the *a* and *b* directions, respectively. The light blue lines denote the optimal fit of |Δ(***r***)| to a composite

function, comprising a cosine term that corresponds to PDW modulations and a quartic term accounting for the underlying inhomogeneity in $\Delta_0$.

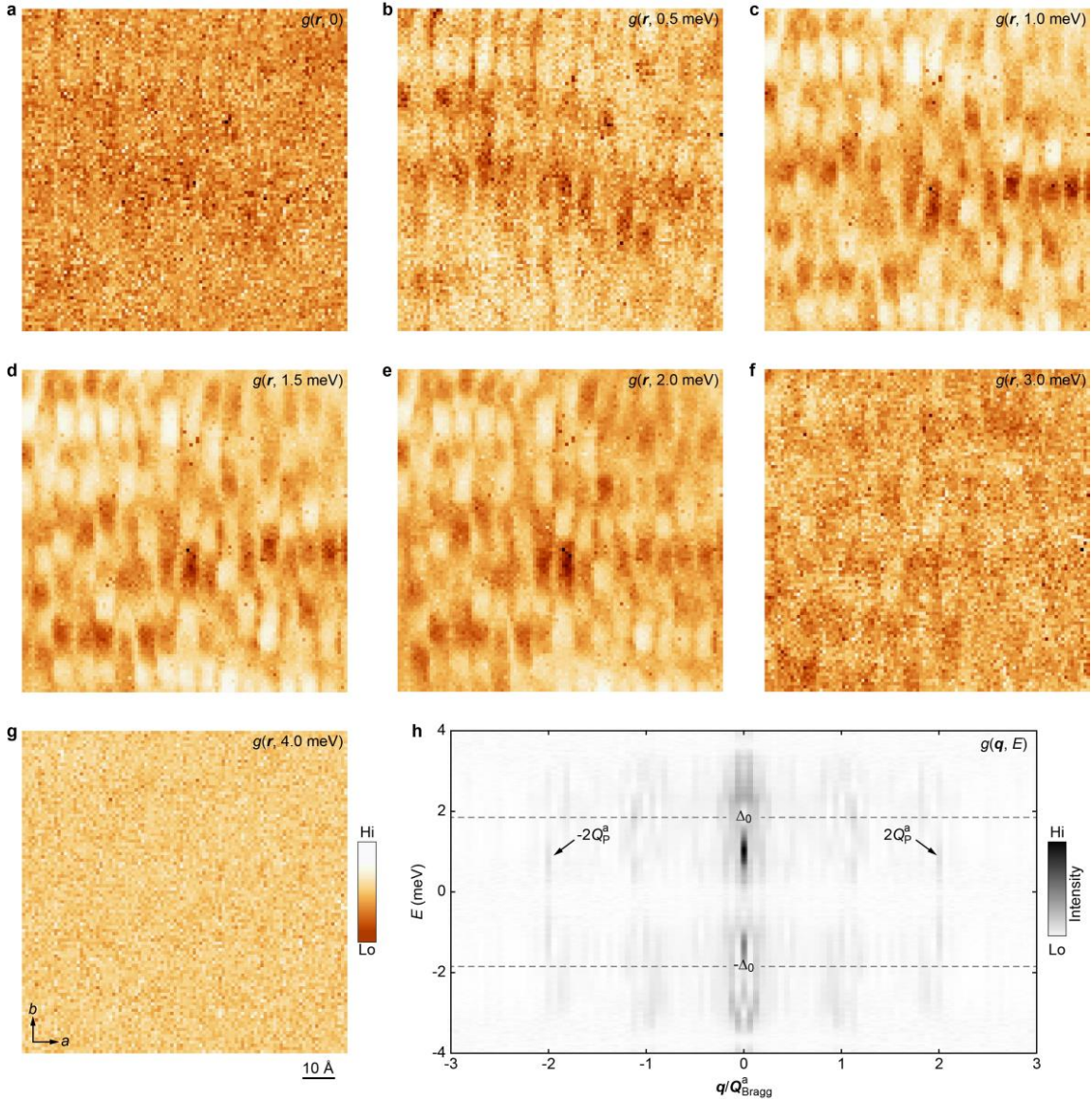

**Extended Data Fig. 4|. Subgap state modulations. a-g**, Mapping low-energy $g(r, E \leq 4$ meV) within the 5 nm × 5 nm FOV, as depicted in Fig. 2a. Intriguingly, the subsidiary CDW modulations, $\rho_{2P_a}$ and $\rho_{P_b}$, can be observed at energies down to $E_F$, but disappear outside the superconducting gaps. **h**, Map of line cuts of the energy-dependent $g(q, E)$ through Γ along the $q_a$ axis at 0.4 K. The two horizontal dashes correspond to the energy positions of the average superconducting gap ($\pm\Delta_0$). It should be emphasized that the half-unit-cell spatial modulations are exclusively resolved around the superconducting gaps.

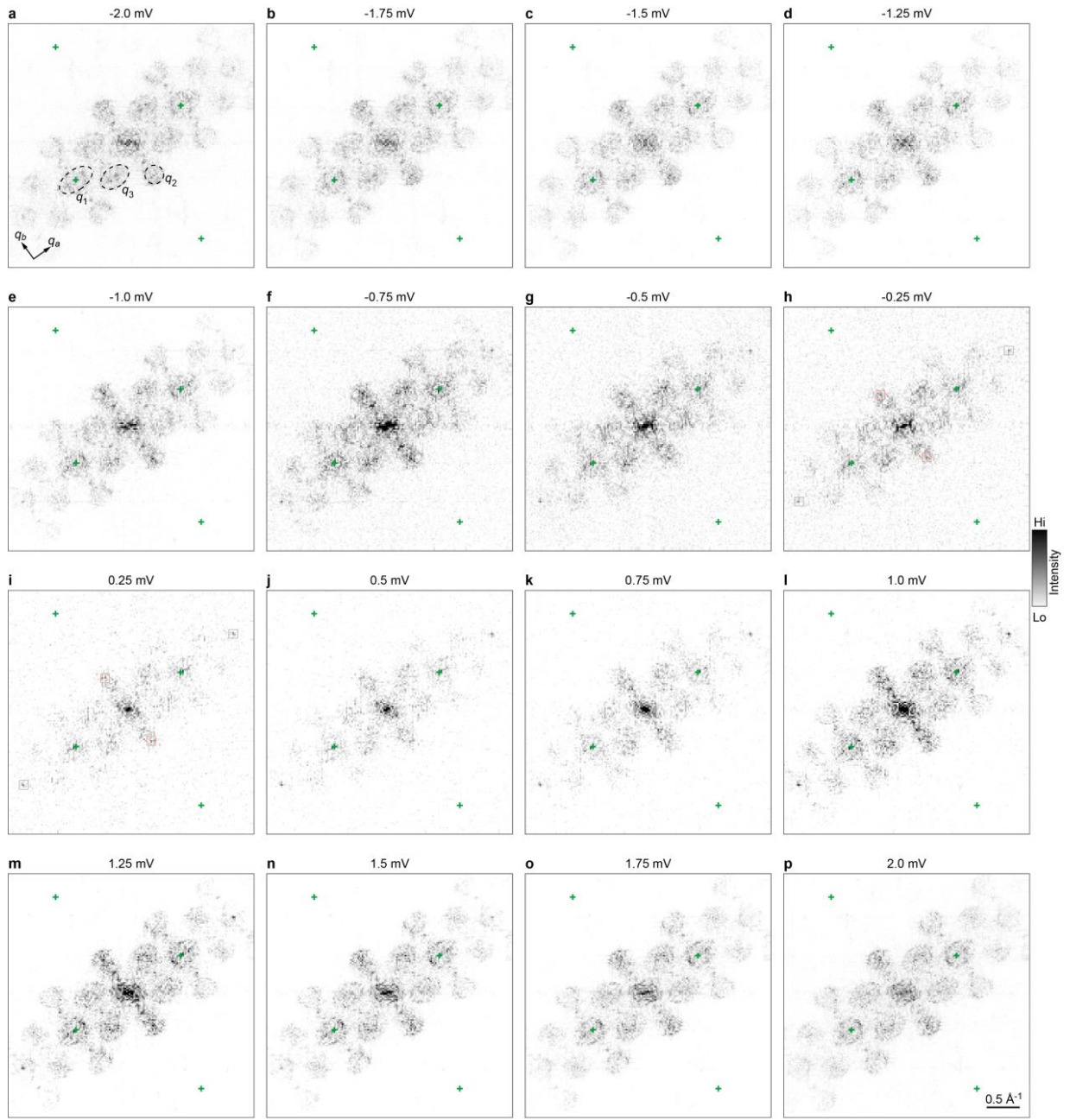

**Extended Data Fig. 5|. Unique observation of subsidiary CDWs within Δ(*r*). a-p**, Dependence of $g(q, E)$ derived from the low-energy $g(r, E)$ maps on the energy $E = eV$ as indicated, measured at 0.4 K and zero magnetic field. Setpoint: $V = 10$ mV, $I = 1$ nA. As the energy approaches $E_F$, unique spatial modulations associated with the bidirectional PDWs emerge at $Q_P^a = (\pm 4\pi/a, 0)$ and $Q_P^b \approx (0, \pm 2\pi/3b)$, highlighted by the squares in **h** and **i**. This unconventional electron pairings substantially suppress the QPI patterns among the electron pockets, as indicated in **a**.

*Supplementary Information*

**Imaging half-unit-cell Cooper-pair density waves in monolayer 1T′-MoTe$_2$**

Fang-Jun Cheng[1], Cong-Cong Lou[1], Ai-Xi Chen[2], Li-Xuan Wei[1], Yu Liu[1], Bo-Yuan Deng[2], Fangsen Li[2], Zi-Qiang Wang[3], Qi-Kun Xue[1,4,5,6,7], Xu-Cun Ma[1,4], Can-Li Song[1,4]

[1]*State Key Laboratory of Low-Dimensional Quantum Physics, Department of Physics, Tsinghua University, Beijing 100084, China*

[2]*Vacuum Interconnected Nanotech Workstation, Suzhou Institute of Nano-Tech and Nano-Bionics, Chinese Academy of Sciences, Suzhou 215123, China*

[3]*Department of Physics, Boston College, Chestnut Hill, MA 0246, USA*

[4]*Frontier Science Center for Quantum Information, Beijing 100084, China*

[5]*Shenzhen Institute for Quantum Science and Engineering and Department of Physics, Southern University of Science and Technology, Shenzhen 518055, China*

[6]*Hefei National Laboratory, Hefei 230088, China*

[7]*Beijing Academy of Quantum Information Sciences, Beijing 100193, China*

## 1. Interfacial engineering of ML 1T′-MoTe$_2$ films

Graphitized SiC(0001) substrates feature a graphene overlayer with adjustable thickness, which can be controlled by varying the annealing temperature and duration under ultra-high vacuum conditions. In this study, the 4$H$-SiC(0001) substrates were all subjected to annealing at an elevated temperature up to 1400°C, resulting in a coexistence of single-layer and multiplayer graphene. This spatial heterogeneity facilitates the realization of diverse electronic properties in the ML 1T′-MoTe$_2$ overlayers grown atop these substrates. As illustrated in Supplementary Fig. 1a, the semi-metallic-like d$I$/d$V$ spectra show two prominent peaks located below and above the $E_F$, designated as $V$1 and $C$1, respectively. Remarkably, the energy positions of these characteristic peaks exhibit spatial variability across different sample regions. For example, from region #1 to region #3, the V1 peak is shifted downwards by 0.17 eV, indicating moderate electron transfer from the underlying substrate to the sample. This downward shift is similarly observed for the $C$1 peak, albeit to a lesser extent. The non-rigid band shifts are most likely attributable to site-dependent epitaxial strain effects. Accompanying the interfacial electron doping, significant modifications in the superconducting energy gap structure are identified. In the slightly electron-doped region (#1, Ref. 36), the superconducting gap features a double-gap structure with a more prominent gap at ± 1.27 meV. However, the moderately electron-doped region (#3) reveals a predominant gap at ± 1.85 meV. Importantly, the superconducting energy gaps in sample region #3 are more well-defined (Fig. 2e,f), with pronounced coherence peaks. This system thus provides us an idea opportunity to investigate the superconducting properties and their spatial dependencies in multi-band superconductors.

A further comparison among various sample regions reveals the unique emergence of electron pockets at the $X$ point of the BZ boundary as the electron doping increases. The emergent electron pockets have two essential consequences. First, they contribute to the stabilization of bidirectional 2 × 5 CDWs by providing additional nesting conditions between the hole pocket at Γ and the electron pockets at $X$ ($Q_C^a$), although the nesting conditions between the hole pocket and the electron pockets along the Γ$Y$ line ($Q_C^b$) become poor to some extent. Second, they prompt three predominant inter(intra)-band scatterings between various electron pockets, labeled as $q_1$, $q_2$ and $q_3$, respectively, as depicted in Extended Data Fig. 1a. These scatterings result in three sets of elliptical or circular scattering rings, which have been clearly identified from the FT images of the energy-varying $g(r, E)$ maps in Extended Data Fig. 1a-d. With increasing energy, the scattering rings gradually enlarge, hinting at the electron-like band dispersions. This can be unambiguously corroborated by extracting the energy-dependent linecuts of $g(\boldsymbol{q}, E)$ along the high-symmetry directions and plotting them in Extended Data Fig. 2e,f.

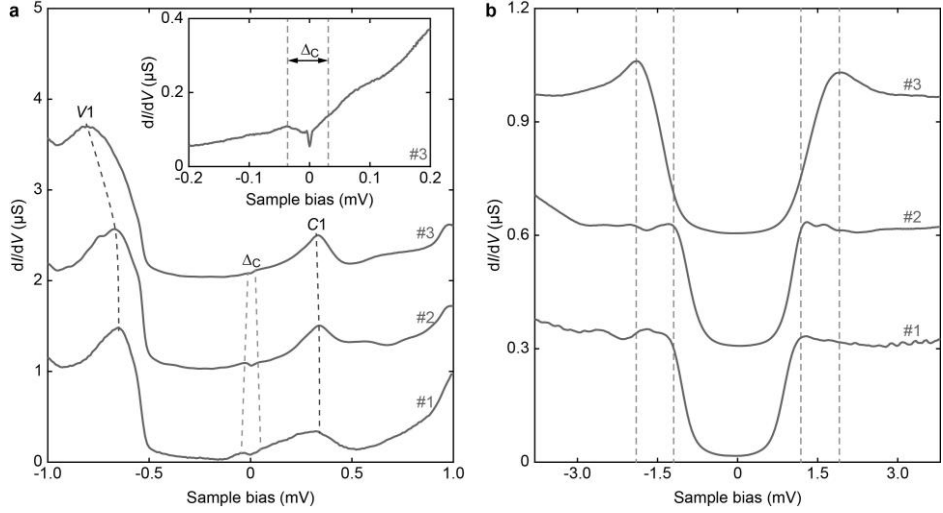

**Supplementary Fig. 1. Interfacial engineering of the electronic structure and superconductivity in ML 1T′-MoTe$_2$. a**, Spatially-averaged differential conductance d$I$/d$V$ spectra across three distinct regions of ML 1T′-MoTe$_2$ on graphitized SiC(0001) substrates. Prominent spectral peaks, denoted as $V$1 and $C$1, are found to shift downwards from sample region #1 to sample region #3, signifying enhanced electron transfer from the electronically heterogeneous graphene substrates. This electron transfer results in a gradual reduction of the CDW gap $\Delta_C$ proximate to $E_F$, as marked by the dashed lines. Inset shows a magnified view of the d$I$/d$V$ spectrum near $E_F$ in region #3. Setpoint: $V = 1.0$ V, $I = 0.5$ nA. **b**, Low-energy-scale d$I$/d$V$ spectra revealing the evolution of superconducting gaps across the corresponding sample regions in **a**. As the electron doping intensifies, the larger superconducting gap (approximately at ± 1.85 meV) becomes dominant, leading to a distinctly defined single gap in sample #3. These findings underscore the important influence of interfacial engineering on the superconducting properties of ML 1T′-MoTe$_2$.

Accompanied by enhanced interfacial electron transfer, the CDW-associated gap size $\Delta_C$ diminishes from 140 meV in sample region #1 to 67 meV in sample region #3, as vividly illustrated Supplementary Fig. 1a. This reduction is consistent with the deteriorating nesting conditions between electron and hole pockets and the fluctuating nature of CDWs in region #3 (Fig. 1d,e and Supplementary Fig. 2 and 3). Furthermore, it is noteworthy that the superconducting gap, appearing on an energy scale of < 2 meV, is substantially smaller than $\Delta_C$, as clearly revealed in the inset of Supplementary Fig. 1a.

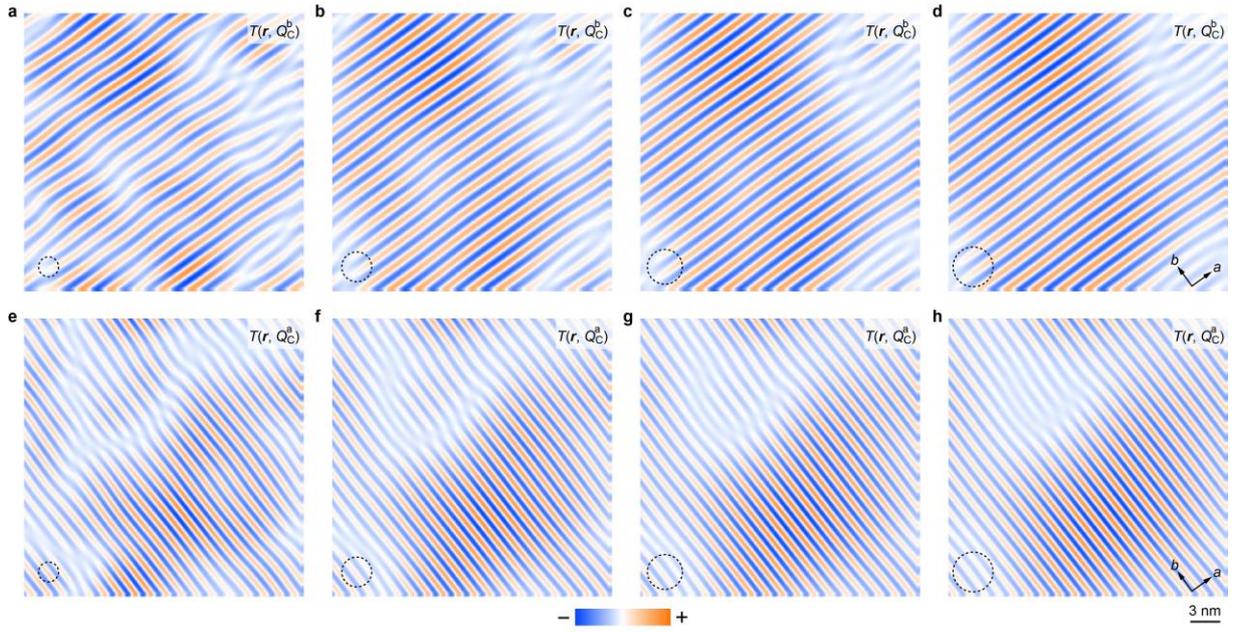

**Supplementary Fig. 2. Fluctuating CDWs. a-h**, Inverse Fourier transforms of the 2 × 5 CDW peaks at $Q_C^a$ and $Q_C^b$, $T(r, Q_C^a)$ (**a-d**) and $T(r, Q_C^b)$ (**e-h**), derived from $T(r, 10\text{ mV})$ in Fig. 1b. The STM topography $T(r)$ is Gaussian-filtered at various cutoff lengths, 20 Å, 30, Å, 35 Å and 40 Å. The dashed circles in the bottom-left corner illustrates the filter size.

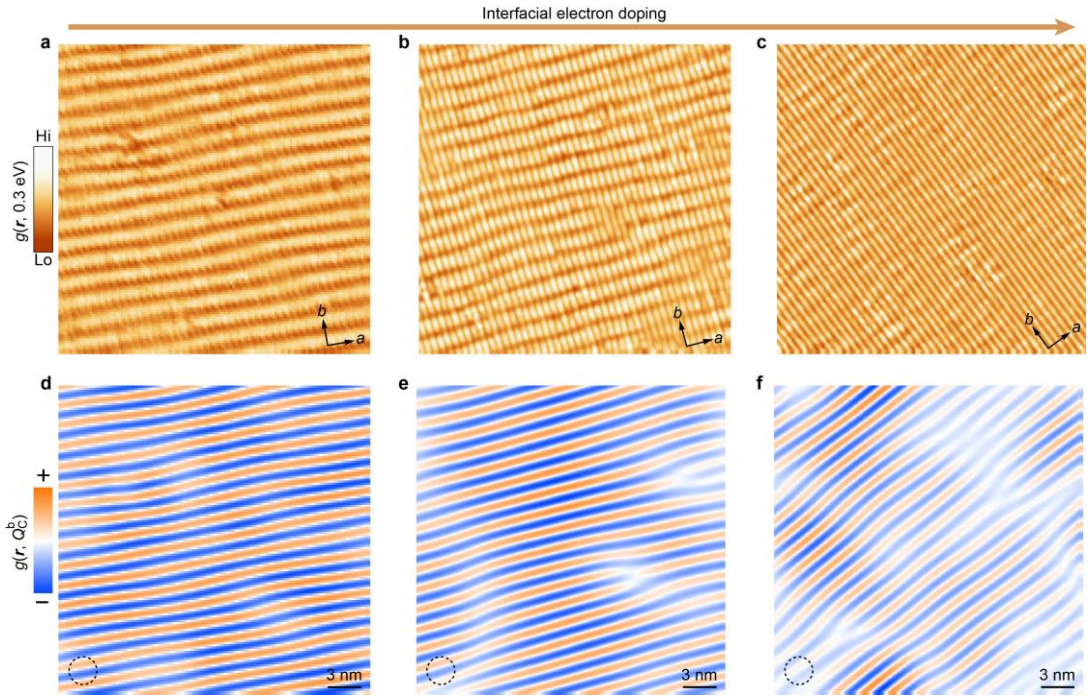

**Supplementary Fig. 3. a-c**, Electron doping dependence of $g(r, E = 0.3\text{ eV})$ maps, showing the evolution of unidirectional CDWs at $Q_C^b \approx (0, \pm 2\pi/5b)$ across various samples regions (#1 to #3, left to right). Setpoint: $V = 0.3\text{ V}$, $I = 1\text{ nA}$. The modulations become increasingly attenuated in region #3, which has the highest

level of electron doping. **d-f**, Inverse Fourier transform of the $Q_C^b$ peaks from Gaussian-filtered $g(r, E)$ maps in **a-c**, revealing spatial CDW modulations with a wavelength of ~ 5$b$. A cutoff length of 25 Å (indicated by dashed circles in the bottom-left corner) is consistently applied for the Gaussian filtering. More remarkably, increasing electron transfer from the substrates leads to the emergence of topological defects.

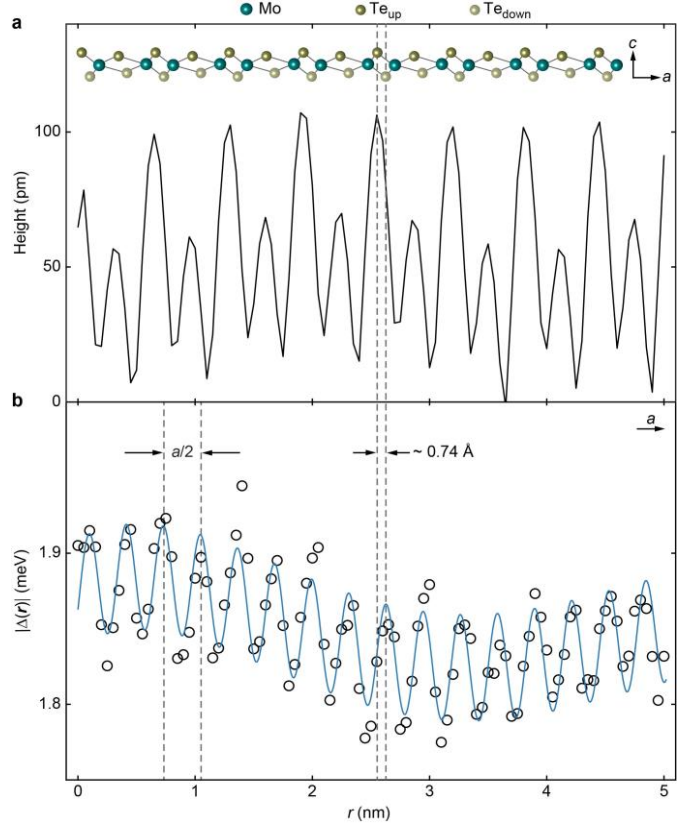

**Supplementary Fig. 4. Atomic-scale correlation between Δ(*r*) maxima and Te sublattices. a,b**, Spatial variations of |Δ(*r*)| and the bulked atomic corrugation from the top Te layer of ML 1T′-MoTe$_2$ along the *a* direction, as derived from the simultaneously acquired STM topography T(*r*) and grid d*I*/d*V* spectra. In **b**, the light blue line represents the optimal fit of |Δ(*r*)| to a composite function comprising both a cosine term and a quartic term. The inset in **a** shows a side view of the atom structure of 1T′-MoTe$_2$. The vertical dashed lines are guides to eye. It is apparent that the two distinctly bulked Te sublattices at the bottom layer exhibit indistinguishable gap maxima.

**2. Time-reversal symmetry breaking in the superconducting state**

In the PDW states, we can simply express the PDWs associated with the intra-band Amperean pairing as $\Delta_P^\alpha(r) = e^{i\theta_\alpha}(\Delta_{Q_P^\alpha} e^{iQ_P^\alpha \cdot r} + \Delta_{-Q_P^\alpha} e^{-iQ_P^\alpha \cdot r})$, where the index $\alpha$ denotes either $a$ or $b$. As thus, the experimentally measured gap magnitude can be read as

$$|\Delta(\mathbf{r})| = |\Delta_0 e^{i\theta_0} + \Delta_P^a(\mathbf{r}) + \Delta_P^b(\mathbf{r})|. \tag{10}$$

Inserting $\Delta_P^\alpha(\mathbf{r}) = e^{i\theta_\alpha}(\Delta_{Q_P^\alpha} e^{iQ_P^\alpha \cdot \mathbf{r}} + \Delta_{-Q_P^\alpha} e^{-iQ_P^\alpha \cdot \mathbf{r}})$ into Eq. 10, it reads as

$$|\Delta(\mathbf{r})| = \sqrt{\begin{array}{l}\Delta_0^2 + 4\Delta_0\Delta_{Q_P^a}\cos(Q_P^a\cdot\mathbf{r})\cos(\theta_a-\theta_0) + 4\Delta_0\Delta_{Q_P^b}\cos(Q_P^b\cdot\mathbf{r})\cos(\theta_b-\theta_0) \\ + 4\Delta_{Q_P^a}\Delta_{Q_P^b}\cos(Q_P^a\cdot\mathbf{r})\cos(Q_P^b\cdot\mathbf{r})\cos(\theta_a-\theta_b) + 4\Delta_{Q_P^a}^2\cos(Q_P^a\cdot\mathbf{r})^2 + 4\Delta_{Q_P^b}^2\cos(Q_P^b\cdot\mathbf{r})^2\end{array}} \tag{11}$$

$$= \sqrt{\begin{array}{l}\Delta_0^2 + 2\Delta_{Q_P^a}^2 + 2\Delta_{Q_P^b}^2 + 4\Delta_0\Delta_{Q_P^a}\cos(Q_P^a\cdot\mathbf{r})\cos(\theta_a-\theta_0) + 4\Delta_0\Delta_{Q_P^b}\cos(Q_P^b\cdot\mathbf{r})\cos(\theta_b-\theta_0) \\ + 4\Delta_{Q_P^a}\Delta_{Q_P^b}\cos(Q_P^a\cdot\mathbf{r})\cos(Q_P^b\cdot\mathbf{r})\cos(\theta_a-\theta_b) + 2\Delta_{Q_P^a}^2\cos(2Q_P^a\cdot\mathbf{r}) + 2\Delta_{Q_P^b}^2\cos(2Q_P^b\cdot\mathbf{r})\end{array}}. \tag{12}$$

Provided that $|\Delta_{Q_P^\alpha}| \ll \Delta_0$ is generally satisfied (Extended Data Fig. 3), we can thus obtain

$$|\Delta(\mathbf{r})| = \Delta_0 + \Delta_{Q_P^a}^2/\Delta_0 + \Delta_{Q_P^b}^2/\Delta_0 + 2\Delta_{Q_P^a}\cos(Q_P^a\cdot\mathbf{r})\cos(\theta_a-\theta_0) + 2\Delta_{Q_P^b}\cos(Q_P^b\cdot\mathbf{r})\cos(\theta_b-\theta_0) +$$

$$2\Delta_{Q_P^a}\Delta_{Q_P^b}/\Delta_0\cos(Q_P^a\cdot\mathbf{r})\cos(Q_P^b\cdot\mathbf{r})\cos(\theta_a-\theta_b) + \Delta_{Q_P^a}^2/\Delta_0\cos(2Q_P^a\cdot\mathbf{r}) + \Delta_{Q_P^b}^2/\Delta_0\cos(2Q_P^b\cdot\mathbf{r}). \tag{13}$$

Experimentally, the intensity of the Fourier peaks in $|\Delta(\mathbf{q})|$ provides a direct measure of the gap modulation amplitude in $\Delta(\mathbf{r})$ at the corresponding wave vector $\mathbf{q}$. Using the measured gap modulations at $2Q_P^a$ i.e. $\Delta_{2Q_P^a} = \Delta_{Q_P^a}^2/\Delta_0 = 0.035$ meV, and $\Delta_0 = 1.85 \pm 0.02$ meV in Fig. 2a, we can estimate $\Delta_{Q_P^a} \approx 0.25$ meV. In Fig. 2c, the ratio between the measured FT intensities at $2Q_P^a$ and $Q_P^a$ is $\Delta_{Q_P^a}/2\Delta_0\cos(\theta_a - \theta_0) \sim 3.54$. Thus, the phase difference between $\theta_a$ and $\theta_0$ is calculated to be $\sim 0.494\pi \cong \pi/2$. This indicates a breaking of time-reversal symmetry in the superconducting state, which well accounts for the more pronounced superconducting gap modulations at $2Q_P^a$ compared to $Q_P^a$. This symmetry breaking can result in our unique observation of half-unit-cell PDWs along the $a$ direction.

In contrast, the spatial superconducting gap modulations at $2Q_P^b$ are negligibly small compared to those at $Q_P^b$ along the $b$ direction, due to the significantly small order parameter $\Delta_{Q_P^b} \approx 0.030$ meV. This finding is in consistent with a zero-phase difference between $\theta_b$ and $\theta_0$. Consequently, there exists a phase difference of $\theta_a - \theta_b = \pi/2$ between the two unidirectional PDW components. This implies that a bidirectional-II PDW ground state has been realized in ML 1T′-MoTe$_2$ (Ref. 3), matching well with our analysis of the subsidiary orders induced by the PDWs.

The coexistence of bidirectional-II superconducting order parameters $\Delta_P^\alpha(\mathbf{r})$ ($\alpha = a$ and $b$) and a uniform order parameter $\Delta_0 e^{i\theta_0}$ inherently results in intricate coupling interactions between them. In the absence of magnetic field, the lowest-order coupling term in the Ginzburg-Landau theory can be expressed as

$$H_c = \beta_1(|\Delta_0 e^{i\theta_0}|^2|\Delta_P^a|^2 + |\Delta_0|^2|\Delta_P^b|^2) + \beta_2[(\Delta_0 e^{i\theta_0})^2(\Delta_P^a \Delta_{-P}^a + \Delta_P^b \Delta_{-P}^b)^* + c.c.] \qquad (14)$$

$$= \beta_1(\Delta_0^2 \Delta_{Q_P^a}^2 + \Delta_0^2 \Delta_{Q_P^b}^2) + 2\beta_2 \Delta_0^2 [\Delta_{Q_P^a}^2 \cos(2\theta_0 - 2\theta_a) + \Delta_{Q_P^b}^2 \cos(2\theta_0 - 2\theta_b)]. \qquad (15)$$

Given the positive $\beta_2$ and $\Delta_{Q_P^a} \gg \Delta_{Q_P^b}$, the lowest free energy is achieved at $\theta_0 - \theta_a = \pi/2$. This condition is consistent with the observed time-reversal symmetry breaking in the ML 1T′-MoTe$_2$ system. A similar mechanism was recently proposed for a kagome superconductor CsV$_3$Sb$_5$ (Ref. 44).

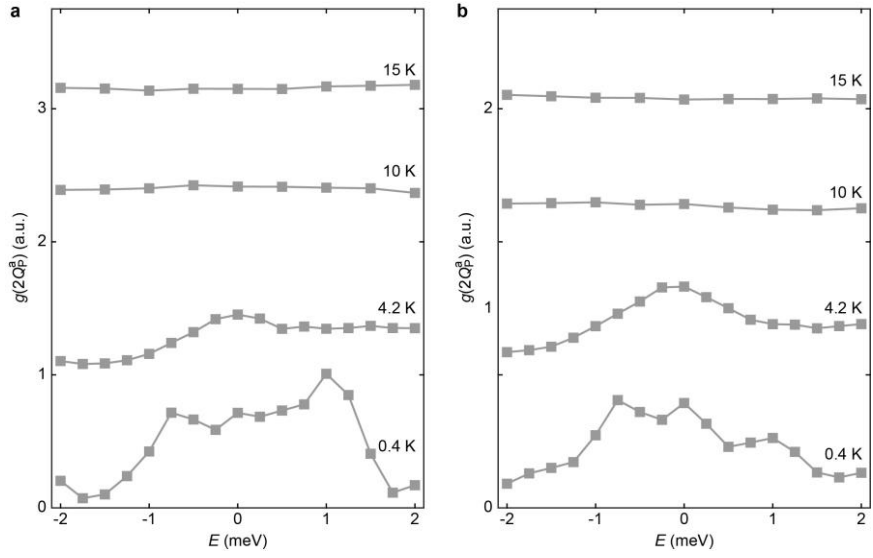

**Supplementary Fig. 5. Temperature-dependent PDW modulations. a**, Extracted $g(2Q_P^a, E)$ intensities as a function of the energy $E = eV$ and temperature. Although the $g(2Q_P^a, E)$ intensities alter little with $E$ in the normal state, they become significantly enhanced around $E_F$ in the superconducting state. **b**, Same to **a** but for the extracted $g(Q_P^b, E)$ intensities.

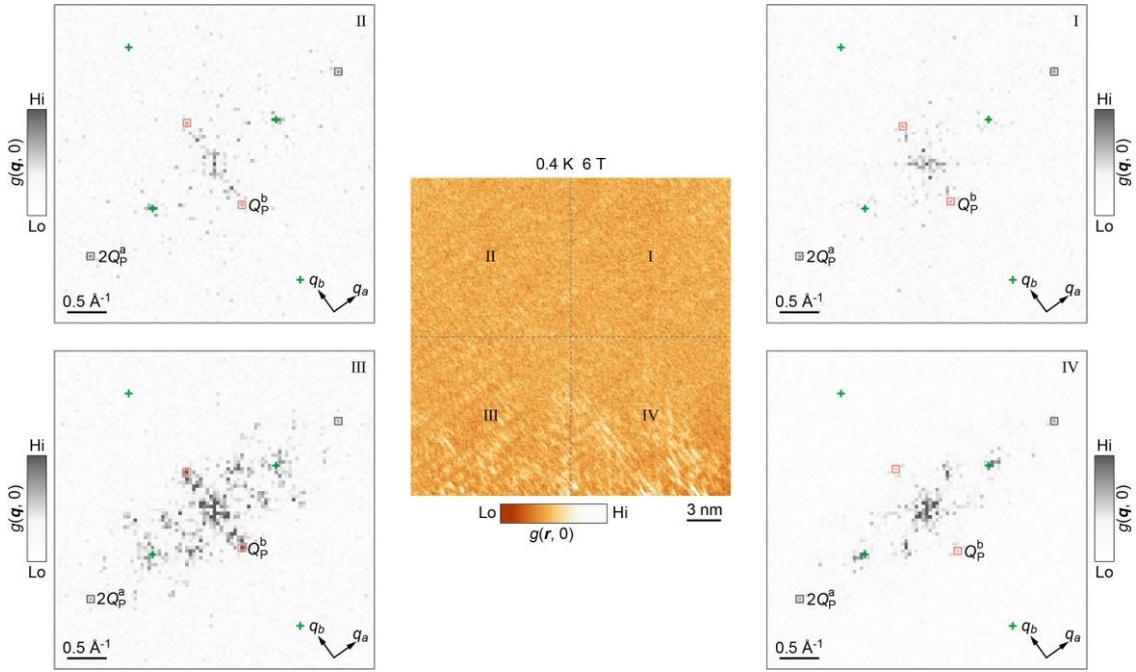

**Supplementary Fig. 6. Spatial robustness of the subsidiary CDWs.** Fourier transform $g(\mathbf{q}, 0)$ from four segmented regions from the zero-bias conductance $g(\mathbf{r}, E = 0)$ map in Fig. 3a. The $g(\mathbf{r}, 0)$ map presents tiny inhomogeneities, specifically with more prominent subgap states at the bottom. Nevertheless, the subsidiary CDW modulations (marked by the squares) essentially remain across the varying regions, as evidenced by the $g(\mathbf{q}, 0)$ maps at the four corners. This indicates the robustness of the subsidiary CDWs, demonstrating their stability and persistence irrespective of tiny local variations in the moderately electron-doped sample.